# Design and construction of the ISR

*Kurt Hübner*

## Introduction

The emergence of the ISR project at CERN is described in the light of the situation at CERN at the end of the 1950s when the CERN Proton Synchrotron (PS) was still under construction. The discussions leading to the project are put into context with world-wide efforts to build larger and more powerful accelerators at that time; the evolution of the project before approval is sketched. The basic design considerations and the most significant technological choices are explained. The construction period is summarized by highlighting important milestones and the performance achieved during commissioning in 1971, the first year of running, is given.

## 1       On the early history of colliders

*First ideas*

The first proposal for colliding beams was made by Rolf Wideröe in a German patent of 1943 which was published only in 1953 due to the circumstances of that time [1]. Figure 1 shows the first page of the patent taken from the scientific biography of Wideröe by Pedro Waloschek [2]. Figure 2 is taken from the patent showing injector and collider. All essential features are described: the counter-rotating particle beams (protons or deuterons) to reach a high reaction energy in the centre-of-mass, a ring-shaped vacuum tube and a magnetic guide field. Even electron-proton reactions were considered. Wideröe discussed the idea with Bruno Touschek who was not very impressed saying that the idea was obvious and trivial [3]. Nobody could see an application of the idea since the reaction rate was simply too low to be of any practical use and, at that time, no scheme for accumulating intense beams was known.

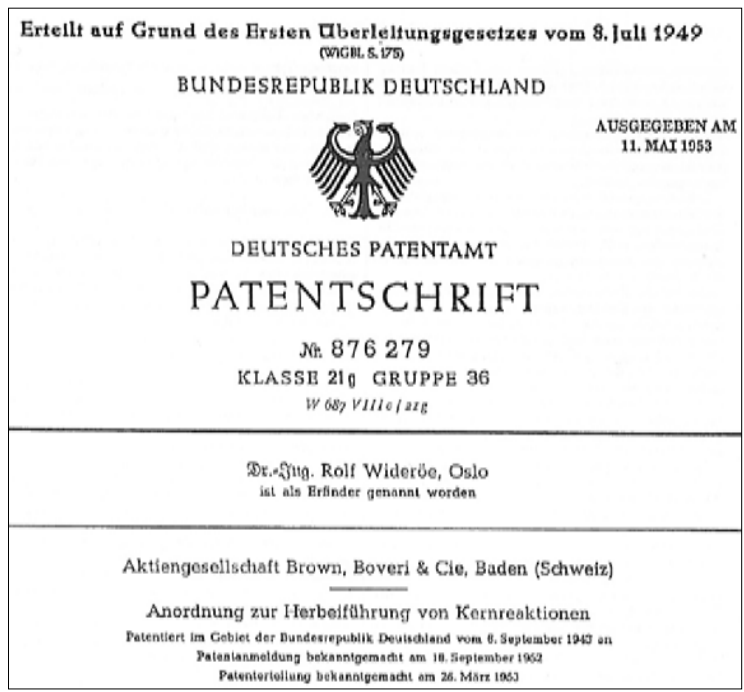

**Fig. 1:** The patent of R.Wideröe introducing colliding beams

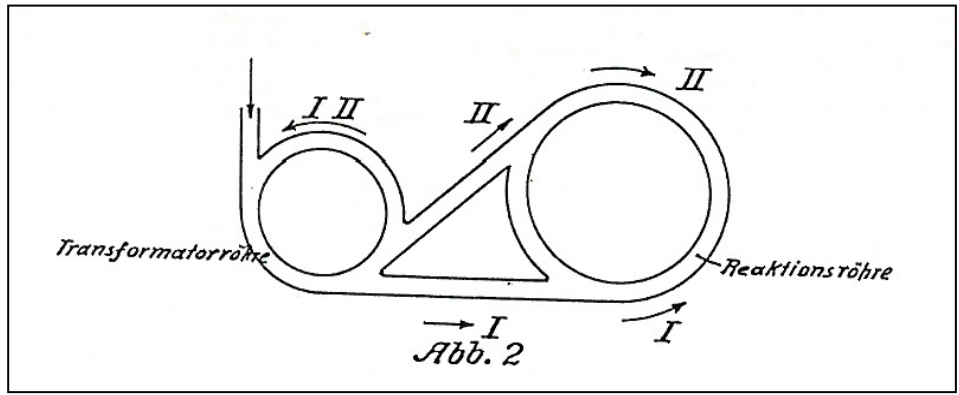

**Fig. 2:** The layout of a collider with injector by R.Wideröe

*Practical ideas*

A breakthrough came with the invention of the so-called radio-frequency (rf) stacking by the MURA Group in the US in 1956 [4] which showed a way to accumulate proton beams of sufficient intensity so that reasonable interaction rates could be hoped for. Based on this technique, this group led by Donald W. Kerst worked out in detail a technical proposal for collisions based on two Fixed-Field Alternating Gradient (FFAG) rings with a common straight section where the beams could collide. The layout is sketched in Fig. 3 [5]. At the same time, Gerard K. O'Neill presented a layout (Fig. 4) with tangential rings having a synchrotron-type magnet structure operating at 3 GeV [6, 7]. Similar ideas occurred to Lichtenberg, Newton, and Ross of MURA [8] and Brobeck [9]. Concentric, intersecting storage rings were suggested later [10], a topology finally adopted for the ISR. The ideas of the MURA Group and of O'Neill were not followed up in the US as cascaded synchrotrons appeared more attractive, leading eventually to the 200 GeV main ring of FNAL. MURA was eventually dissolved [11] and O'Neill became interested in electron-electron collisions.

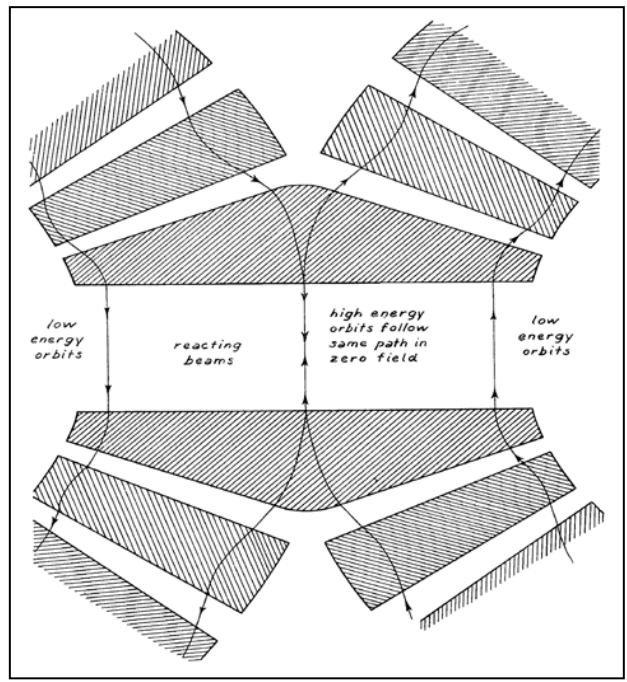

**Fig. 3:** Layout of a collider proposed by the MURA Group [5]

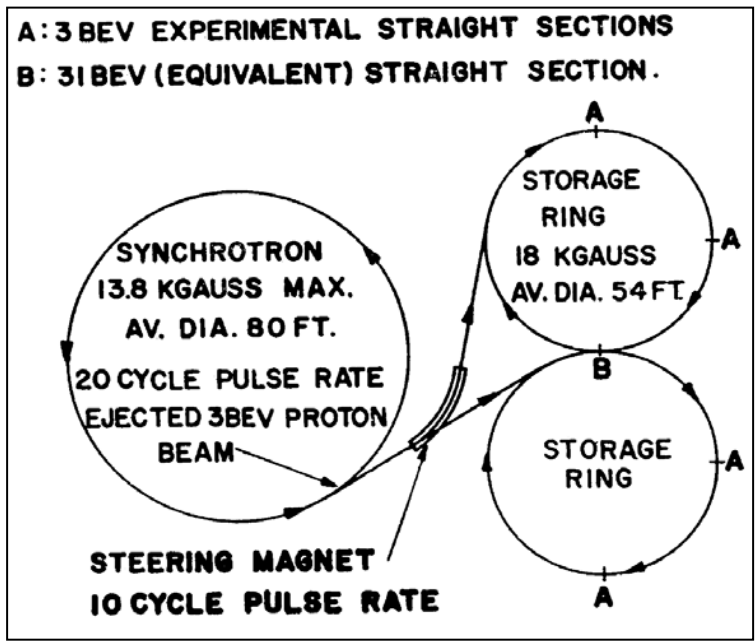

**Fig. 4:** Proton storage rings fed by a synchrotron [6]

*Electron–electron and electron–positron colliders*

In order to illustrate the general context of the CERN ISR studies, it is useful to recall the parallel activities in the field of $e^-e^-$ and the $e^-e^+$ colliders. The design of the Princeton–Stanford $e^-e^-$ rings with a beam energy of 500 MeV started in 1957 and the rings operated from 1961 onwards in Stanford. Their operation revealed the strong effect of synchrotron radiation on the vacuum system which has been an important issue for electron and positron rings ever since [12]. Also in 1957 the design of the $e^-e^-$ VEP-1 with 160 MeV per beam started at the Kurchatov Institute in Moscow under the leadership of Gersh Budker. VEP-1 operated from 1965 at BINP in Novosibirsk [13].

The lineage of single rings for counter-rotating $e^-e^+$ beams, which culminated in LEP at CERN, was initiated at Frascati by Touschek with ADA [14] designed for beams with an energy of 200 MeV. Amazingly, it took less than a year from proposal to first operation in 1961. Since the beam power of the Frascati 1 GeV electron synchrotron turned out to be insufficient for adequate positron production, ADA was moved in 1962 to the more powerful 1 GeV electron linac of LAL at Orsay where real physics experimentation started. ADA was followed by VEPP-2 in 1964 at BINP and by ACO in 1965 at LAL. Their beam energies were already much higher, 0.7 GeV and 0.5 GeV, respectively.

## 2  The emergence of the ISR

In 1956, still during the construction of the CERN Proton Synchrotron (PS), the CERN Council established the Accelerator Research (AR) Group to be led by Arnold Schoch following a proposal of John Adams. This group was expanded in 1959 with manpower that became available after the PS construction had been terminated. It initially studied plasma acceleration and an electron collider with 100 MeV beam energy in a FFAG ring, but in 1960 interest swung to proton–proton storage rings fed by the PS. A proposal of tangential rings was made in December 1960 and, in 1961, it was decided that the AR Division, formed at the beginning of 1961, should study the proton storage rings and a large 300 GeV synchrotron. In 1962, Intersecting Storage Rings were proposed, thus considerably simplifying the project.

In order to obtain a reasonable luminosity in the ISR, accumulation of the beam injected from the PS was imperative. The method of choice was rf stacking invented and promoted by MURA [4, 11]. Given the importance of the performance of this new method, an experimental proof was indispensable. Hence the idea of constructing a small electron ring to test this method came up in 1960 and the ring, the CERN Electron Storage and Accumulation Ring (CESAR), was ready in 1964 (Fig. 5). It had a circumference of 24 m. To mimic the proton behaviour at 25 GeV in the ISR correctly, synchrotron radiation had to be negligible but v/c close to 1. This led to the choice of an electron energy of 2 MeV and a low magnetic bending field of 130 G. This low field led to trouble in operation as, in addition, solid bending magnets had been chosen. Their substantial residual magnet field compared to the nominal field of 130 G was difficult to control. Nevertheless CESAR quickly demonstrated rf stacking [15] (Fig. 6) giving welcome momentum to the ISR project.

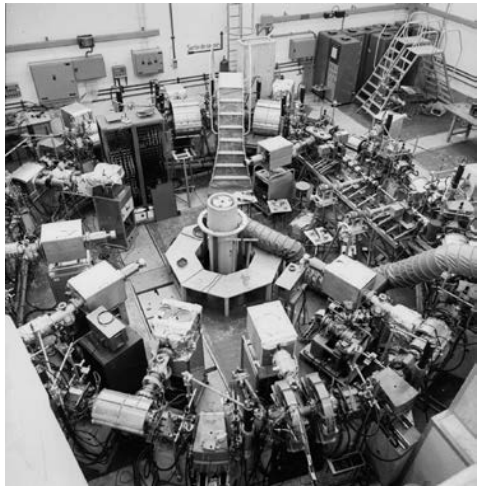

**Fig.5:** The CERN Electron Storage and Accumulation Ring (CESAR)

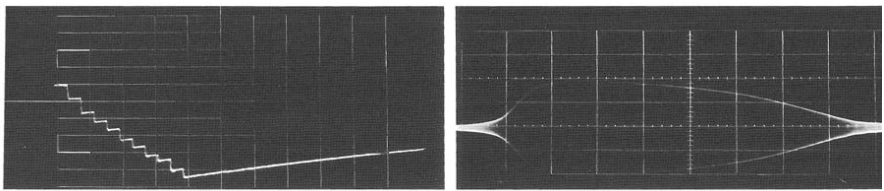

**Fig. 6:** a) Accumulated electron beam current as a function of time in CESAR;  b) Particle density of the stack versus particle momentum

In order to channel the discussions, an ECFA Working Group chaired by E. Amaldi was formed in 1963 which recommended the ISR and the 300 GeV synchrotron. However, the Homeric debate went on between those who favoured a facility to peep at interactions at the highest energies and those who preferred intense secondary beams with energies higher than the PS could provide. Those against the ISR were also afraid of the leap in accelerator physics and technology required by this venture, which appeared to them as a shot in the dark. In May 1964, the AR Division presented the ISR design report [16] and in November the design report of the 300 GeV accelerator [17]. The following year, CERN Council approved the ISR as a supplementary programme in June and then approved the project in December with K. Johnson as project leader after the financing had been clarified. The prevailing argument had been "to remain competitive for as low a cost as possible" given that the ISR was estimated to be much cheaper than a 300 GeV synchrotron. The cost of the former was estimated to be 312 MCHF [16] compared to 1556 MCHF [17] for the latter. A very detailed account of the period up to the decision can be found elsewhere [18].

## 3 ISR design

No small-scale proton collider had ever been built before, hence no extrapolation was possible. The only experience at CERN with an accelerator of that size was the PS. A number of leading team members had indeed acquired their expertise during PS design, construction and running-in, which turned out to be very beneficial for the project, in particular for its rapid and uneventful construction, though one might argue today that this stifled somewhat the quest for new solutions.

*Magnet lattice*

The magnet lattice requirements were different from the PS: long straight sections were needed at the crossing points to make space for the experiments and the horizontal aperture for the beam had to be larger as the rf stacking required a large momentum bite. The long straight sections were inserted between two focusing magnets in order to minimize the disturbance to the beta-functions and to provide a small vertical beam size at the crossing point as the luminosity is inversely proportional to the vertical beam size. Matched low-beta insertions with vanishing dispersion, common today in all colliders, had not yet been invented.

Three alternative types of magnet lattice had been considered: a separated-function lattice where the magnets have either bending or focusing function; combined-function lattices either of FODO or FOFDOD type consisting of magnets with dipole and quadrupole fields providing both bending and focusing. Since elaborate poleface-windings were foreseen because they are easier to implement in combined-function magnets, this type of magnet was preferred but also for easier access to the very demanding vacuum system. The latter argument led also to the choice of FODO because the access to the FD junction in FOFDOD is not so easy, as experience in the PS had shown. An additional argument for the combined-function lattice was the claim that a separated-function lattice would increase the cost up to 1.7 MCHF. Table 1 gives a synopsis of the parameter ranges considered, the final choice and the consideration leading to the decision.

**Table 1:** Considerations and choices for the ISR lattice

|  | Range | Chosen | Consideration |
|---|---|---|---|
| Interaction regions (No.) | 6–8 | 8 | Avoid betatron stop-bands |
| Betatron oscillations per turn $Q$ (h/v) | 6–9 | 8.8/8.7 | $nQ = p(N/2)$, $N/2$ – number superperiods |
| Lattice periods | 45–60 | 48 | Betatron phase advance between $\pi/4$ to $\pi/3$ |
| Half-periods in outer arc | 14–24 | 16 | Limit on circumference |
| Half-periods in inner arc | 4–12 | 8 | Geometry |
| Full crossing angle | 9–32° | 14.77° | Numerology relative to PS |

The resulting topology with the transfer lines relative to the PS is shown in Fig. 7. The ISR circumference was chosen to be 1.5 times that of the PS resulting in 942.64 m (300 $\pi$ m). Inspection of Fig. 8 showing one octant reveals indeed a clear FODO structure in the outer arc but it is harder to clearly determine the type of lattice in the inner arc.

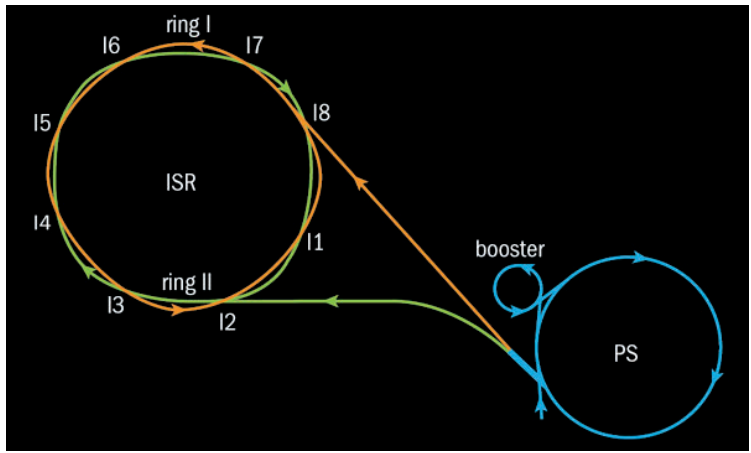

**Fig.7:** Schematic layout of the ISR, its transfer lines and its injector, the PS

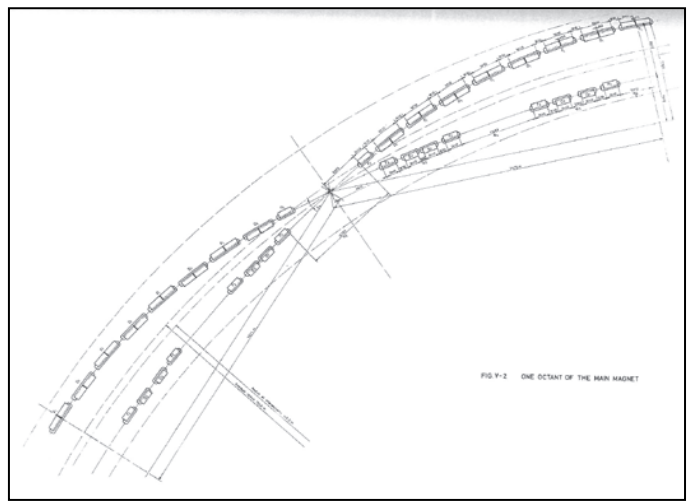

**Fig.8:** One octant of the ISR magnet lattice [16]

The ISR were argued for initially as an improved facility for the CPS also providing additional fixed-target beams of high-intensity but low duty cycle. A token of this is the layout shown in the design report [16] (see Fig.1 of P. Bryant's contribution) which still features a hall for fixed-target experiments in the lower right corner but this hall disappeared very quickly from the drawings during the construction stage. The extraction channel towards the West Hall, however, was constructed but never equipped.

*Magnets*

The magnets were designed for protons of 28 GeV/*c*, the maximum the PS could supply. The nominal bending field was 1.2 T implying a bending radius of 78.6 m. Each ring contained 60 long magnet units and 72 short units. The long units (L= 5.03 m) were made up of two blocks and the short units (*L* = 2.44 m) of one block. The 32-turn coil was made of copper. Figure 9 gives the magnet cross-section displaying also the pole-face windings and vacuum chamber which can be accessed easily including its heating elements (not shown) for the bake-out in situ. A photo of a magnet is shown in Fig. 3 of P. Bryant's contribution. In the best tradition of CERN, where key elements of an earlier accelerator are often used for mundane purposes by the next accelerator, the bending magnets are still in service but as elements of the beam dump of the LHC as can be seen in Fig. 10. A number of auxiliary magnets completed the magnet system.

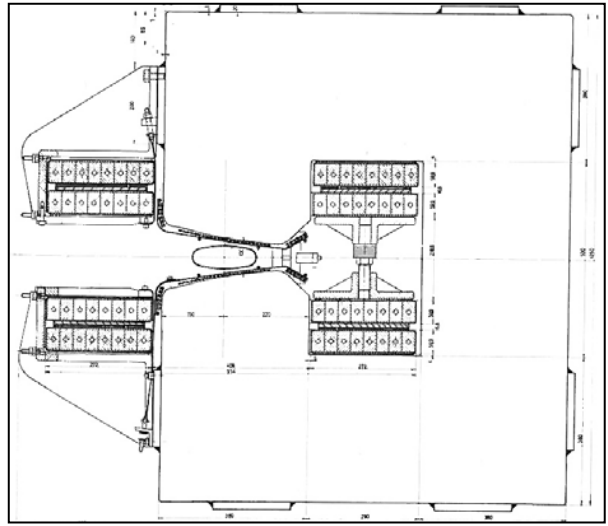

**Fig.9:** Cross-section of an ISR combined-function magnet [19]

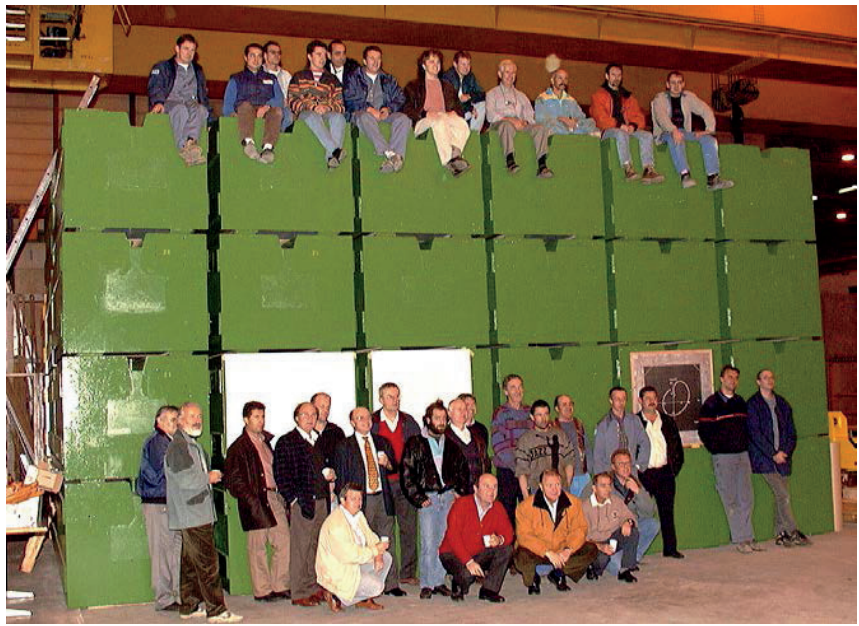

**Fig.10:** Cores of ISR bending magnets as component of the LHC beam dump

Given the bending radius, the energy loss per turn by synchrotron radiation can be calculated. It was $6 \cdot 10^{-14}$ GeV, which is indeed very small compared to 28 GeV. Hence, no radiation damping would fight the beam blow-up by non-linear resonances and by the beam-beam effect as in electron accelerators. Since this was new territory, it fired the fear that the ISR might never work. However, this eventually turned out to be a chimera.

*Vacuum system*

The ISR key performance parameter, the integrated luminosity, is proportional to

$$\int (I_1 \cdot I_2 / h_{\text{eff}}) \, dt$$

with all three variables depending on time $t$. The currents $I_i$ of the counter-rotating beams decay due to nuclear and single-Coulomb scattering, and the effective beam height $h_{eff}$ gets blown up by multiple-Coulomb scattering of the protons on the residual gas. Hence, an ultra-high vacuum system was imperative for the performance of the ISR, in order to achieve a reasonable beam lifetime and to limit the beam blow-up as a function of time. Imposing a beam loss of less than 50% and a growth of $h_{eff}$ of less than 40% in 12 h, implying a drop to not less than 18% in luminosity after 12 h, leads to a requirement that the pressure be less than $10^{-9}$ Torr ($N_2$ equivalent) averaged around the circumference. The pressure in the interaction regions had to be less than $10^{-11}$ Torr to limit the background for the experiments. To produce such an ultra-high vacuum system extending over a total length of nearly 2 km was one of the biggest technological challenges of the project.

CESAR had been a valuable test bed to guide the choice of the vacuum technology: a stainless-steel vacuum chamber of low magnetic permeability and bakeable in situ to 300°C (initially baked only to 200°C); flanges with metal seals; sputter ion pumps (350 l/s) complemented with Ti-sublimation pumps (2000 l/s) in critical places. Figure 9 indicates the position of the vacuum chamber in the bending magnet. The long vacuum chambers in the interaction points were particularly challenging as they had to be designed with a minimum of mechanical support and with very thin walls to reduce the loss and scattering of secondary particles produced in the collision point. Engineering highlights were the self-supporting chambers with 0.3 mm wall-thickness made from Ti and with 0.2 mm thickness made from stainless steel. INCONEL of 0.2 mm thickness was also used [20]. Figure 11 shows an example.

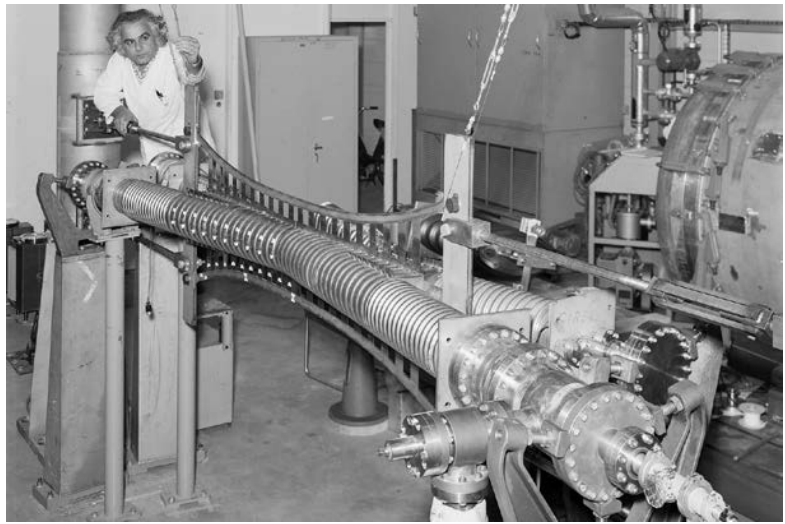

**Fig.11:** A thin-wall ISR vacuum chamber for an intersection region

Clearing electrodes inside the vacuum chambers were foreseen to remove the electrons created by ionization of the residual gas and accumulating in the potential well of the d.c. proton beam. Damping resistors reduced the quality factor of the electromagnetic eigen-modes in all cavity-like chambers to prevent collective instabilities of the beams.

## 4    ISR construction and commissioning [21]

*ISR tunnel*

The tunnel was built using the cut-and-fill method implying excavation and removal of more than one million cubic metres, mainly moraine material, since the tunnel was 15 m wide and it had to be put on competent rock, i.e., molasse in the Geneva basin. Figure 12 illustrates this point. The tunnel floor was

12 m above the level of the PS to reduce the amount of material to be removed. The erection of the tunnel proceeded so as to protect the foundations of the magnets as much as possible. First, two concentric concrete footings (1 m deep, 1.5 m wide) were laid. The floor in between was left 20 cm higher than the final general level. The footings formed the base for the mobile crane and, then, for the foundations for the pre-fabricated concrete main walls of the tunnel. Once a section of the tunnel had been completed, the remaining molasse was excavated to the general level and trenches were prepared for the concrete beams supporting the magnets [19].The cross-section of the ring tunnel is shown in Fig. 13. Its height was 6.5 m and 4.0 m under the hook of the crane. The finished tunnel and the concrete support beams for the magnets can be seen in Fig. 14 showing also the first magnet in the tunnel used for a positioning trial.

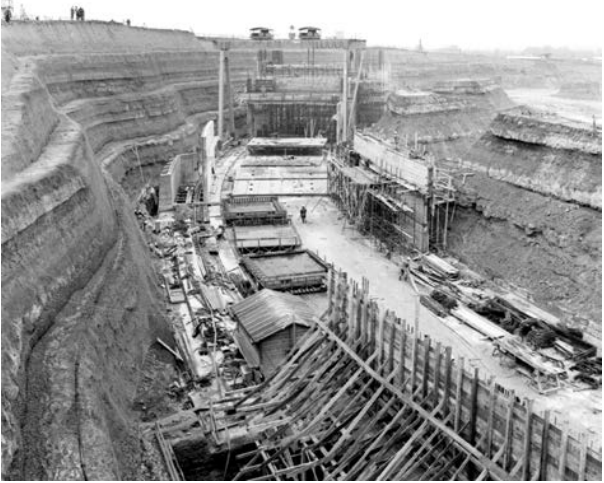

**Fig.12:** The excavation of the ISR tunnel

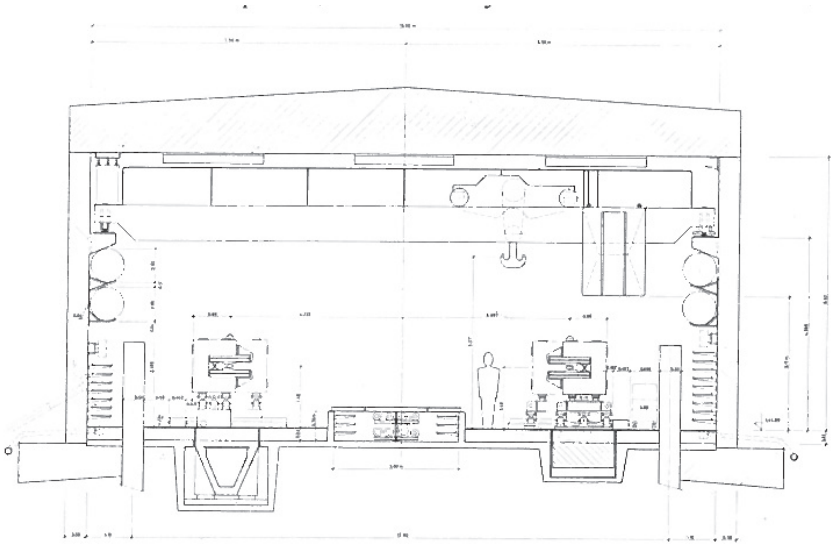

**Fig.13:** The cross-section of the ISR tunnel [21]

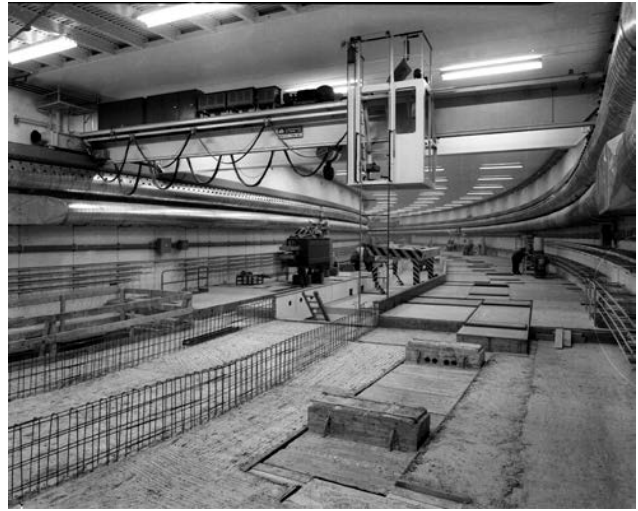
**Fig.14:** A view of the ISR tunnel

*Construction milestones*

The ISR construction got off to a flying start. Excavation started in November 1966, less than 12 months after approval. The pre-fabricated structure of the tunnel was in place in 1969 and installation in six octants had started. By 1967, all major magnets had been ordered. Two prototype magnets were delivered in early 1968 and measured. The bulk production of magnet steel started at the beginning of 1968 and the whole order (11 kt) was delivered by October. The West Hall became available for the assembly, testing, and storage of the magnets in 1969 (Fig. 15). The final race started in 1970. In April, the transfer lines from the PS were ready for tests with beam and the last ring magnet was installed in May. The earth shielding was complete in July and ring 1 was ready for injection in October.

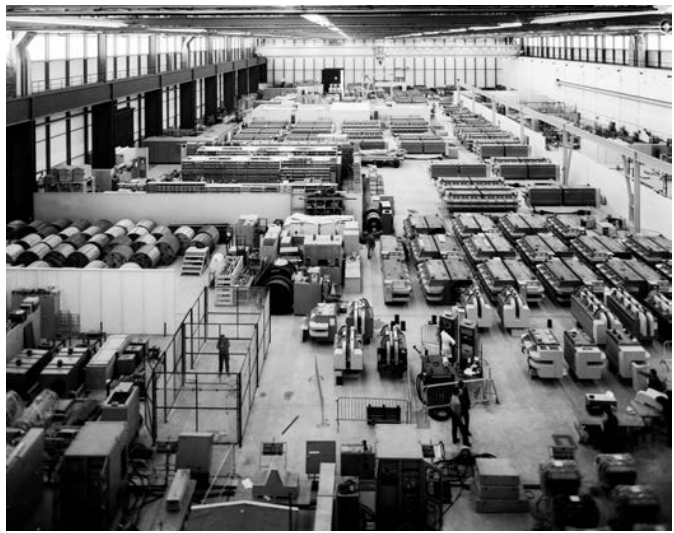
**Fig. 15:** Assembly, testing and storage of ISR magnets in the West Hall

*Commissioning*

The first 15 GeV/*c* proton beam was injected into ring 1 on 29 October 1970 and a circulating beam was quickly obtained. The uncorrected closed-orbit distortions were about 20 mm peak-to-peak in the horizontal plane and 8 mm in the vertical plane, which could easily be corrected. The number of betatron oscillations per turn was as expected indicating correct focusing of the ring. A first trial of beam accumulation by rf stacking led to 0.65 A showing a satisfactory efficiency of 70% (longitudinal phase space density of the stored beam over that of the injected beam) and confirming the findings in CESAR. An example of early stacking is shown in Fig. 16.

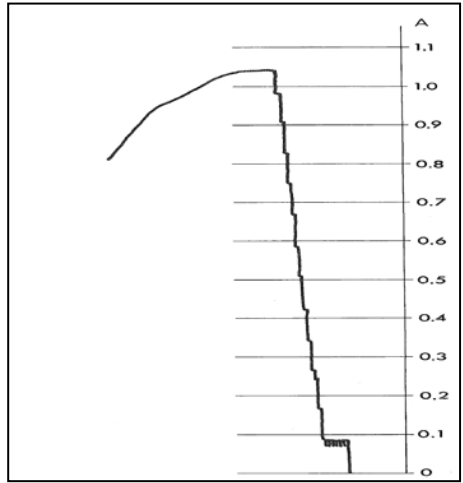

**Fig.16:** Build-up of proton current as a function of time (going to the left) during rf-stacking

In January 1971, the second ring became available and first proton–proton collisions were recorded on 27 January, anxiously observed by the team to see whether the beam–beam effect would quickly destroy the beams as some simulations had predicted. However, nothing catastrophic happened and, to general relief, the beam decay was as expected from the measured vacuum pressure. Regular physics runs started in February with 15 GeV/$c$ beams, and collisions at 26.5 GeV/$c$, the maximum scheduled for the PS, were obtained in May, providing a centre-of-mass energy equivalent to a 1500 GeV proton beam on a fixed target. The beam currents were gradually increased during the year reaching 10 A at the end as illustrated on Fig. 17. The maximum luminosity obtained in 1971 was $3 \cdot 10^{29}$ cm$^{-2}$ s$^{-1}$, a quite respectable performance when compared with the design luminosity of $4 \cdot 10^{30}$ cm$^{-2}$ s$^{-1}$. The beam decay rate was less than 1% per hour at currents of 6 A in both rings, much better than the design value of less than 6% per hour. The ISR operated 1800 h in its first year (800 h for colliding beam physics) with a remarkable availability of 95%.

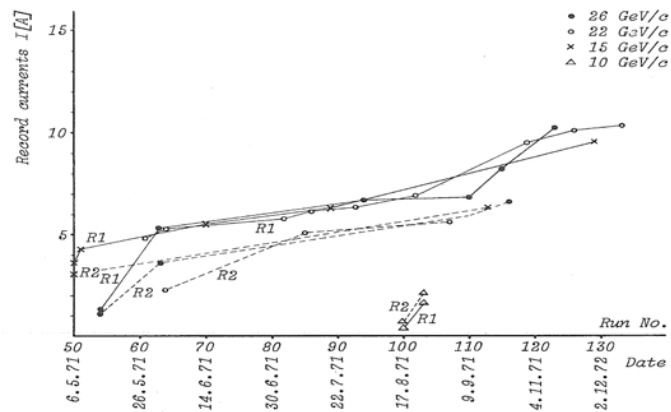

**Fig.17:** The evolution of the stacked proton beam current in the first year of operation for ring 1 and 2 with beam momentum as parameter

The successful completion of the project, terminated officially on 1 March 1971 within budget (332 MCHF in 1965 prices), was duly celebrated in an inauguration ceremony on 16 October where the photograph (Fig. 18) of two of the main-players, Eduardo Amaldi and Kjell Johnsen, was taken. The third one was Viktor Weisskopf who had the stamina to push their vision until it was accepted by the funding authorities catapulting CERN to the very high-energy front. It would turn out that it was not quite enough to be at this frontier since this state was eventually not fully exploited by CERN.

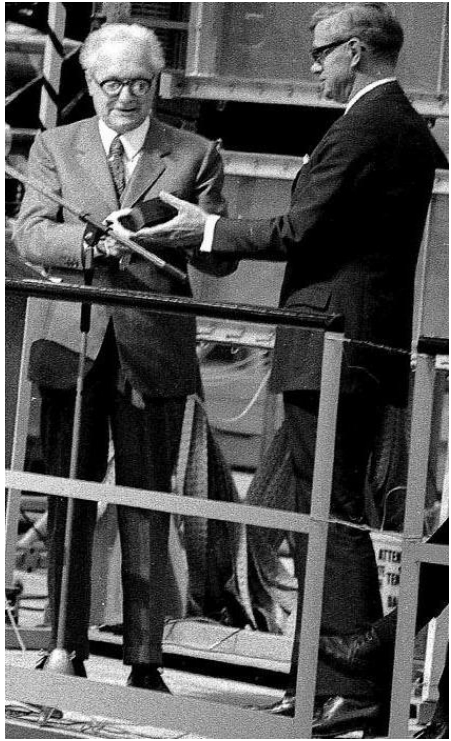

**Fig.18:** Eduardo Amaldi and Kjell Johnsen at the ISR inauguration ceremony in October 1971

## 5   Conclusions

The ISR construction went very smoothly due to a careful and meticulous preparation by a competent, dedicated team which designed and constructed the conventional components as well as possible, knowing that this provides the best basis for dealing later with unknowns which might appear. Some accused the team of overdesign and waste of resources. However, this careful approach provided the potential for the later gradual, but spectacular improvement in performance until the ISR were decommissioned as a collider in 1983. A token of this is the fact, that the ISR luminosity record was not broken until 1991 when an $e^-e^+$ collider, CESR in Cornell, took over.

The ISR was a solid basis for the development of all the hadron colliders to come such as the proton–antiproton collider in the CERN SPS and at FNAL, the p–p and ion–ion collider at BNL and, eventually, LHC at CERN. It was a fine and unique instrument, or as Weisskopf put it at the closure ceremony in 1984: *"First considered a window into the future, it turned out to be more"*.